\documentclass[runningheads]{llncs}
\usepackage[T1]{fontenc}
\usepackage{graphicx}
\usepackage{multirow}
\usepackage{url}
\begin{document}
\title{Whole-Body Lesion Segmentation in 18F-FDG PET/CT} 
%

\author{Jia Zhang\inst{1*} \and
Yukun Huang\inst{1*} \and
Zheng Zhang\inst{1} \and
Yuhang Shi\inst{1} }
%
%
\institute{United Imaging Healthcare, Shanghai, China \\
\email{yuhang.shi@united-imaging.com} \\
{*} These authors contributed equally to this work.}
\maketitle              
\begin{abstract}

\end{abstract}

\section{Abstract}

There has been growing research interest in using deep learning based method to achieve fully automated segmentation of lesion in Positron emission tomography–computed tomography(PET-CT) scans for the prognosis of various cancers. Recent advances in the medical image segmentation shows the nnUNET is feasible for diverse tasks. However, lesion segmentation in the PET images is not straightforward, because lesion and physiological uptake has similar distribution patterns. The Distinction of them requires extra structural information in the CT images. The present paper introduces a nnUNet based method for the lesion segmentation task. The proposed model is designed on the basis of the joint 2D and 3D nnUNET architecture to predict lesions across the whole body. It allows for automated segmentation of potential lesions. We evaluate the proposed method in the context of AutoPet Challenge, which measures the lesion segmentation performance in the metrics of dice score, false-positive volume and false-negative volume. The code is available at \url{https://github.com/jiazhang93/segmentation}

\section{Introduction}
F-18 fluoro-2-deoxyglucose (F-18 FDG) radiopharmaceutical PET glucose metabolism was recognized as a biomarker of tumor for risk stratification or prognostic of cancers. Extraction of the potential tumor regions and calculation of the derivatives, such as metabolic tumor volume and total lesion glycolysis, allows for quantitative analysis of the cancer status. The widely recognized method of potential tumor segmentation is the thresholding method, such as, setting SUV thresholding values to segment the potential tumor regions\cite{ref14}. Thanks to the recent development in deep learning based medical image segmentation tasks, fully automated methods were explored in the extraction of potential lesions in PET images \cite{ref18,ref19,ref20}. Despite the advances in methods on medical image segmentation, tumor lesion segmentation on the whole body is still a challenging task. As not only the lesions express high FDG uptake, but also some healthy organs, such as brain, bladder, and kidney, show high FDG update, which called physiological uptake. 

Among all the deep learning models, the nnUNet \cite{ref_nnuent}, a UNet based framework, achieved the best performance of the Medical Segmentation Decathlon challenge in 2018. The nnUNet automatically adapts the UNet structure and configures to datasets of different segmentation tasks with extensively explored preprocessing, training, inference and post-processing steps. In this work, we investigated to use the state of-art nnUNet to predict lesions in the whole-body PET/CT images. Further, we designed a joint 2D-3D model based on the nnUNet to further improve the lesion prediction accuracy.

Our main focus of this work is to develop automated methods to ensure the accurate lesion detection and segmentation, as well as avoiding the potential false positives. We extended the nnUNet for lesion segmentation for whole body FDG-PET/CT. In this work, our main contributions were: 1) implementing ResNet18 as the backbone for the nnUNet, which achieved better performance compared to the original UNet as the backbone; 2) combing 2D and 3D nnUNet to capture both information on large field of view on axial axis and detailed information of 3D direction for further performance improvement; 3) proposing a post-processing method to further reduce false positive values. 

The proposed methods were tested in the context of AutoPet Challenge and the corresponding lesion segmentation performance was evaluated in the metrics of dice score, false-positive volume and false-negative volume.  



\section{Methods}

\subsection{Dataset Description}

\begin{figure}[htb]
\includegraphics[width=\textwidth]{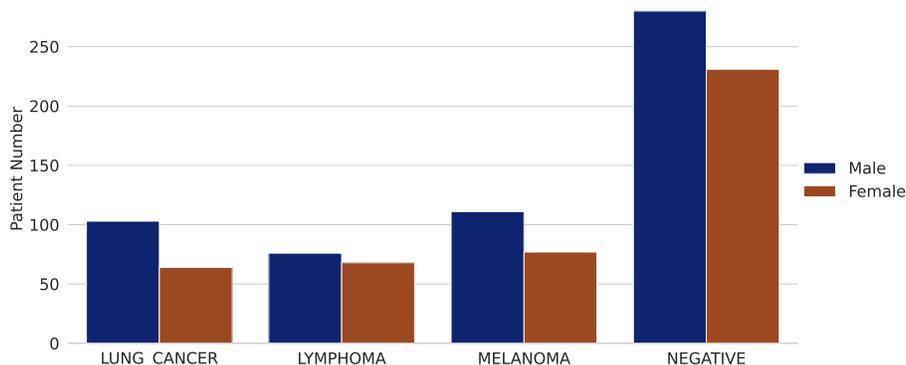}
\caption{The distribution of cancer types among patients in the autoPET database.  } \label{fig1_example_figs}
\end{figure}

The training dataset used for developing our models were acquired from the 2022 AutoPet Challenge. The challenge consists of patients with historically proven malignant melanoma, lymphoma, lung cancer and negative control patients. A total of 900 patients underwent FDG-PET/CT examination resulted in 1014 PET/CT studies. The potential lesions are manually labelled by two experienced clincal experts.

\subsection{Model Design}

\begin{figure}[htb]
\includegraphics[width=\textwidth]{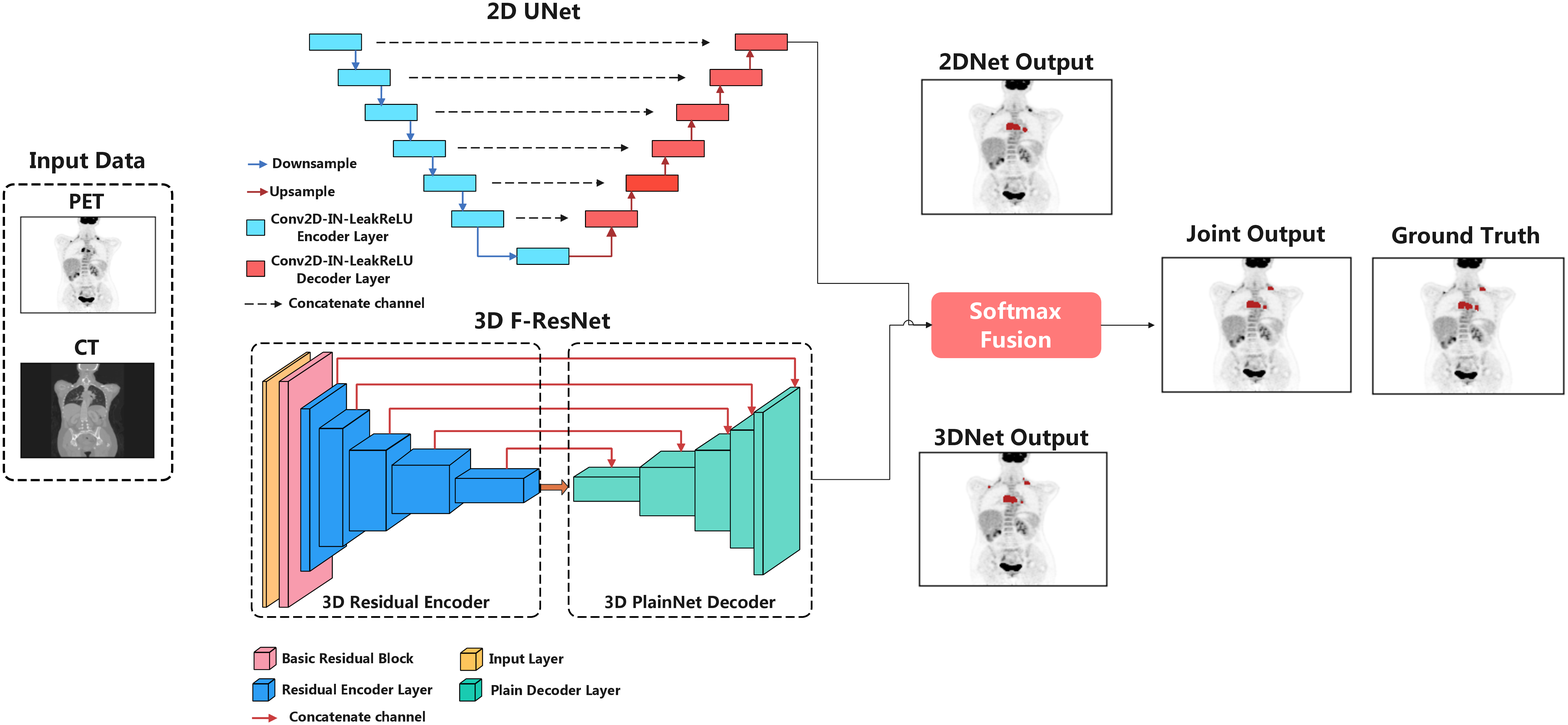}
\caption{The Joint 2D-3D model} \label{fig1_model}
\end{figure}

The performance of different backbones of nnUNet were compared, including 3D UNet (baseline), 2D UNet, 3D low-resolution ResUNet (3D L-Res), 3D full-resolution ResUNet (3D F-Res), cascade 3D ResUNet (Cas-Res) and the joint combination of the 2D UNet and the 3D F-Res models (Joint).

     \subsubsection{3D F-Res model}
    To create 3D F-Res model, the ResNet18 was integrated into the encoder of UNet. The residual units allow deeper network connection and facilitate better information propagation in the training procedure \cite{ref_atten_resunet}.

    \subsubsection{Cas-Res model}
    The Cas-Res model consists of two stages. In the first stage, the 3D L-Res model was trained on the downsampled low-resolution PET/CT images. The predicted outputs were then resampled to the original size of the images. In the second stage, the resampled low-resolution prediction was set as the third channel input to 3D F-Res model for training. 
    
    \subsubsection{Joint 2D-3D model}
    The PET images are relatively low in resolution. Therefore, some lesions are only visible in one image plane. The 2D model are more sensitive to these lesions, compared to the 3D model. To improve the detection accuracy, we combined the prediction results of the 2D UNet model and the 3D F-Res model. The softmax outputs of 2D and 3D models were weighted to produce the final prediction (Equation \ref{equ_softmax}), where $\alpha$ was set to 0.55 to give higher weight to the 3D output. 
    
    \begin{equation}
       softmax_{pred} = \alpha \times softmax_{3D} + (1 - \alpha) \times softmax_{2D}
    \end{equation}\label{equ_softmax}
    

\subsection{Training}

All models are trained from scratch with a five-fold cross-validation method. 

In the data preparation stage, the studies were randomly stratified and splitted into 5 equal sized subsets, using the disease type and number of lesions as the strata. Four of the five subsets were retained as the training data for training the model. The remaining one subset was used as validation data. 

The aforementioned models were trained following the training process specified in the nnUNet paper. Differently, the following settings were adopted in our use case: the model received the original PET images and the resampled CT images as two input channels and took the lesion masks as the output, the batch size was set to 2; the initial learning rate was 0.01 and updated under a polynomial schedule; the stochastic gradient descent (SGD) optimizer was used with a momentum of 0.95; the loss function was the sum of dice loss and cross entropy loss; the models were trained up to 1300 epochs; the input patch size of the 2D U-Net was set to 448×448; the input patch size of the 3D U-Net was set to 128×128x128. Due to the limited computational time required by the autoPET challenge, for the 3D models, only two best-performing folds out of five were used.

The cross-validation process was repeated 5 times by setting different subset as the validation data.

\subsection{Post-processing}
The predicted lesions with the region of interest smaller than 4 voxels were were removed. As the extreme slices of PET images are noisy, we removed the connected components on the three layers located at the bottom of the PET images. 

\subsection{Performance Evaluation}
We evaluated the models on the test dataset using the following 3 metrics provided from the challenge, 1) foreground dice score of the lesions, 2) false positive volume, which is the volume of the false positive connected components that do not overlap with the ground truth positives, and 3) false negative volume, which is the volume of the positive connected components in the ground truth that do not overlap with the predicted lesion mask.

We internally ranked our models by applying weights of the above mentioned metrics (dice score: 0.5, false positive volume: 0.25, and false negative volume: 0.25) and selected the best ranked model for submission to the AutoPET challenge. 



\section{Results}

\subsection{Performance Comparison}
We compared the performance of the above mentioned models on patients with melanoma, lung cancer, lymphoma and negative cases. The fold 1 and 2 were selected for the 3D full resolution ResUnet in the Joint model. Overall, the Joint model achieved the best ranking among all other models, followed by the 3D full resolution ResUNet (3D F-Res) model, the 2D UNet, the 3D low resolution ResUNet (3D L-Res), the Cascade ResUNet (Cas-Res), and the 3D UNet baseline (Table \ref{tab_result_performance}). The dice scores of the Joint model were the highest of all patient types. 
\begin{table}  \setlength{\tabcolsep}{1mm} \centering
\caption{Performance comparison of different models on disease types. Baseline was the 3D UNet, 3D L-Res denoted the 3D low resolution ResUNet, 3D R-Res denoted the 3D full resolution ResUNet, Cas-Res denoted the cascade ResUNet, Joint denoted the joint 2D and 3D ResUNet model, DSC denoted the foreground dice score, FPV denoted the false positive volume, FNV denoted the false negative volume. The bold numbers indicated the best performed metrics across all models.}\label{tab_result_performance}
\begin{tabular}{c|c|c|c|c|c|c} 
\hline

{\bfseries Metrics} &  {\bfseries Baseline} & {\bfseries 2D UNet} & {\bfseries 3D L-Res} & {\bfseries 3D F-Res} & {\bfseries Cas-Res} & {\bfseries Joint}  \\ 
\hline

Melanoma (DSC)    &  0.62      & 0.67            & 0.66      & 0.73      & 0.69      & \bfseries{0.78}  \\    
Melanoma (FPV)    & 9.46      & 1.12            & \bfseries{0.83}      & 4.81      & 5.00      & 1.98 \\
Melanoma (FNV)    &  31.53      & 8.86           &  12.51     &\bfseries{4.08}      & 6.23     & 5.69 \\ 
Lung Cancer (DSC) &  0.70      & 0.80           & 0.77      & 0.82      & 0.81      & \bfseries{0.83} \\ 
Lung Cancer (FPV) &  10.45      & 1.57          &  \bfseries{0.82}     & 5.64      & 4.24     & 1.60 \\ 
Lung Cancer (FNV) &   19.09     & 11.16         & 14.11      & \bfseries{4.20}      & 8.98      & 5.40 \\ 
Lymphoma (DSC)  &  0.68         & 0.72          & 0.71      & 0.77      & 0.74      & \bfseries{0.78} \\ 
Lymphoma (FPV)  &  10.70        & 3.60          & \bfseries{2.29}      & 8.60      & 8.66     & 3.36 \\ 
Lymphoma (FNV)  &  6.47         & 8.21         & 9.93      & \bfseries{3.38}      & 6.92       & 3.89 \\ 
Negative (FPV)  &  11.97        & 1.28          & \bfseries{0.51}      & 5.76     & 5.16       & 2.46 \\ 
Total (DSC)      &  0.67        & 0.72          & 0.71      & 0.77      & 0.74      & \bfseries{0.79} \\
Total (FPV)      &  11.07        & 1.63          & \bfseries{0.87}      & 5.97      &5.48       & 2.36  \\
Total (FNV)     & 9.93        & 4.60          & 6.05      &  \bfseries{1.93}     & 3.61      & 2.50 \\
Model ranking (\#)    & 6        & 3          & 4      &  2     & 5      & 1\\
\hline
\end{tabular}
\end{table}
The dice score of lung cancer from the Joint model was the best (0.83) compared to that of melanoma (0.78) and that of lung cancer (0.78). The total FNV for the 3D F-Res was the lowest (1.93) and relatively high for the 2D UNet (4.60), while, the total FPV for the 3D F-Res was high (5.97) and relatively low for the 2D UNet (1.63). The FPV and FNV of the joint model were between that of the 2D and 3D models.

The 3D L-Res model showed the best false positive volume (FPV), while remained high false negative volume (FNV), indicating large number of missing lesions. The 3D F-Res model resulted in the the lowest FNV for all patient types. 


%

\begin{figure} [!htb]
\includegraphics[width=\textwidth]{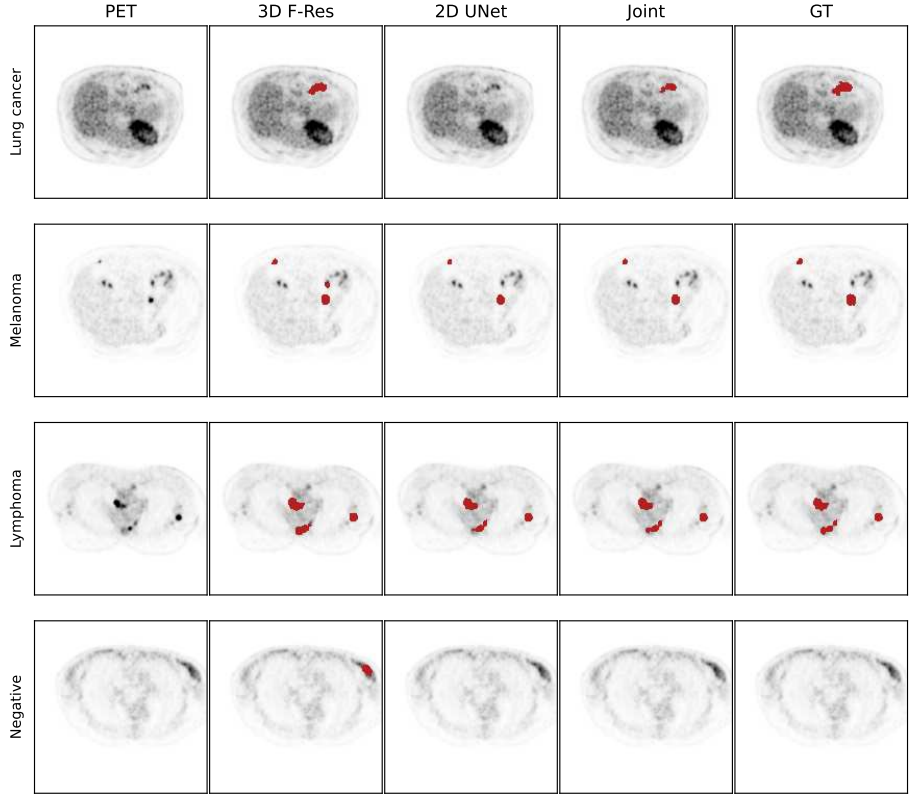}
\caption{Examples of lesion segmentation results from 2D UNet, 3D ResUnet, and joint model compared to ground truth label. Patients with lung cancer, lymphoma, melanoma, and negative patients were selected as examples.}
\label{fig_pred_example}
\end{figure}

The Figure \ref{fig_pred_example} showed the corresponding prediction results of 2D UNet, 3D F-Res and Joint model for four cases of melanoma, lung cancer, lymphoma, and negative category compared to the ground truth label in axial axis. The missed lesion of the lung cancer from the 2D UNet model was compensated by the 3D F-Res model. The false positives in the melanoma case and negative case predicted by the 3D F-Res model was eliminated by the 2D UNet.

\begin{figure}[!htb]
\includegraphics[width=\textwidth]{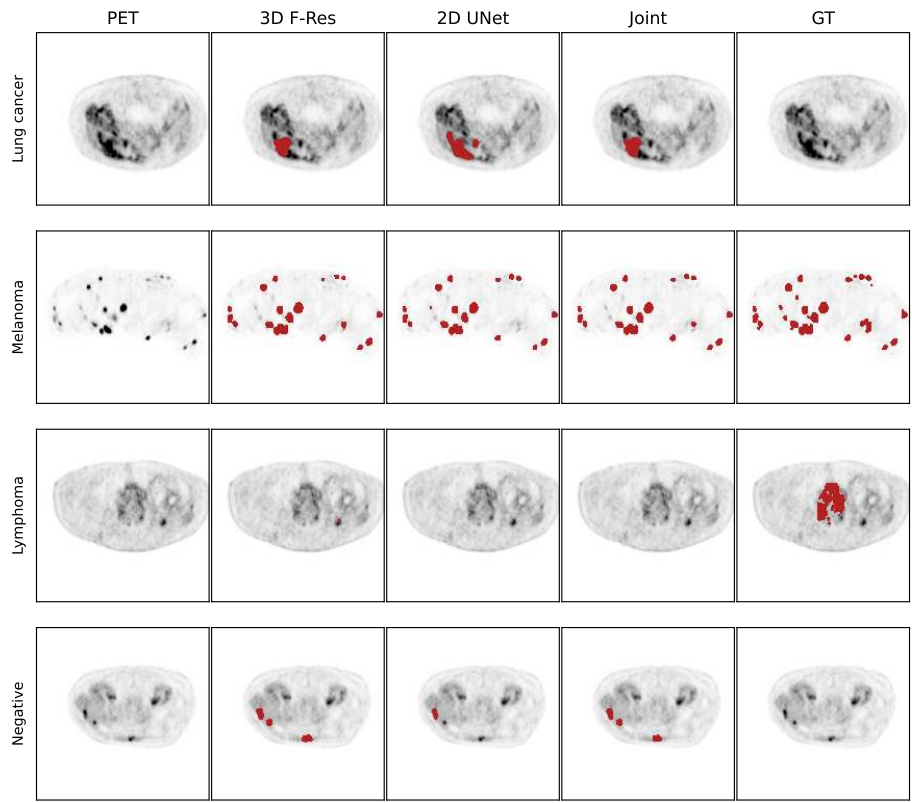}
\caption{Difficult cases of maximum intensity projection of FDG-PET/CT images processed by 2D, 3D and joint 2D-3D models.} \label{fig_difficult_cases}
\end{figure}

Despite the overall good performance, some difficult cases were challenging for the models to make accurate prediction (Figure \ref{fig_difficult_cases}). Physiological uptake on intestine area from the Lung cancer case was falsely identified by all three models. The models were not able to detect all lesions and there existed false positives for the melanoma cases with high tumor burden, where the metastasis could appear in the skin and internal organs. In Lymphoma and negative cases, there were false positives predicted by all models.


\section{Discussion and Conclusion}
We have presented a joint whole-body lesion segmentation approach on FDG-PET/CT images. The Joint model combined the 2D UNet and 3D ResUNet based on the nnUNet framework. The proposed Joint model achieved the best performance ranking compared to all other methods on the randomly splitted validation set from the AutoPet Challenge 2022.

The Joint model balanced the FPV and FNV of the 2D and 3D models and still improved the dice score of all disease categories. The 3D F-Res held the second place on our internal model ranking, and showed the best FNV of all disease categories. The patch size of the 3D model was limited by the GPU memory and could not capture enough global information for the whole-body lesion segmentation application. In combination with the 2D model, which showed better performance on FPV and covered larger region of interest on the axial axis, the 3D model was thus complemented and resulted in a balanced Joint model (Figure \ref{fig_pred_example}). 

The 3D model with ResUNet backbone structure showed better performance compared to the baseline 3D model with UNet backbone structure. The deeper network and residual units of the ResUNet18 contributed to propagate better information through the network \cite{ref_atten_resunet}.  

The performance of the nnUNet was degraded by cascading the prediction results from the low-resolution model. The low resolution model covered more body region while sacrificed the image resolution thus missed detailed information. This was reflected from the best FPV but high FNV of the model performance (Table \ref{tab_result_performance}), meaning that a large proportion was missed by the 3D L-Res model. The cascade model improved the FNV by incorporating the 3D F-Res model compared to the 3D L-Res model, however it did not show improved performance compared to the 3D F-Res model. The low resolution model result may not be suitable to act as a direct input channel in the cascade connection.

The Joint model achieved better performance in segmenting suspicious lesion in cases with lung cancer, compared to those with melanoma and lymphoma. Despite the noises from manual labeling, this may due to the high tumor burden in these two lesion types and the uncertainty of melanoma metastasis through different part of the human body. For further work, lesion segmentation on specific disease type may help to improve the model performance.

\end{document}